\documentclass[nofootinbib,aps,prb,reprint,amsmath,amssymb,groupedaddress]{revtex4-2}

\usepackage{graphicx}
\usepackage{dcolumn}
\usepackage{bm}
\usepackage{color}
\usepackage{hyperref}
\usepackage{braket}

\newcommand{\ve}[1]{\mathbf{#1}}
\newcommand{\ud}{\,\mathrm{d}}
\newcommand{\e}{\,\mathrm{e}}	
\newcommand{\ci}{\,\mathrm{i}}
\newcommand{\exciting}{{\usefont{T1}{lmtt}{b}{n}exciting}}

\begin{document}

\title{Maximally localized Wannier functions within the (L)APW+LO method}

\author{Sebastian Tillack}
\affiliation{Institut f\"ur Physik and IRIS Adlershof, Humboldt-Universit\"at zu Berlin, Berlin, Germany}
\email{sebastian.tillack@physik.hu-berlin.de}
\author{Andris Gulans}
\affiliation{Institut f\"ur Physik and IRIS Adlershof, Humboldt-Universit\"at zu Berlin, Berlin, Germany}
\author{Claudia Draxl}
\affiliation{Institut f\"ur Physik and IRIS Adlershof, Humboldt-Universit\"at zu Berlin, Berlin, Germany}

\date{\today}

\begin{abstract}
We present a robust algorithm that computes (maximally localized) Wannier functions (WFs) without the need of providing an initial guess. Instead, a suitable starting point is constructed automatically from so-called local orbitals which are fundamental building blocks of the basis set within (linearized) augmented planewave methods. Our approach is applied to a vast variety of materials such as metals, bulk and low-dimensional semiconductors, and complex inorganic-organic hybrid interfaces. For the interpolation of electronic single-particle energies, an accuracy in the meV~range can be easily achieved. We exemplify the capabilities of our method by the calculation of the joint density of states in aluminum, (generalized) Kohn--Sham and quasi-particle band structures in various semiconductors, and the electronic structure of $\beta$-Ga$_2$O$_3$, including electron and hole effective masses.
\end{abstract}


\maketitle

\section{\label{Introduction}Introduction}

In the past two decades, maximally localized Wannier functions (MLWFs) became a well established tool in solid state calculations. Due to their localized nature they are superior to the equivalent Bloch representation in terms of chemical interpretation. They provide inexpensive access to both single-particle eigenvalues and eigenfunctions at any point in reciprocal space in terms of the so called Wannier interpolation scheme. Typically, the starting point for the calculation of MLWFs for a quantum mechanical system is a set of single-particle Kohn--Sham (KS) wave functions obtained from density-functional theory (DFT). The nowadays most commonly used approach to MLWFs in solids is based on works by \citet{Marzari1997} (MV) and \citet{Souza2001}. Given a set of single-particle orbitals, the MV algorithm approaches a set of MLWFs by an iterative minimization of the target functional $\Omega$, measuring the spread of the Wannier functions (WFs). In general, this optimization problem is non-linear and high-dimensional. Consequently, the result may strongly depend on the starting point for the minimization, and the algorithm can be easily trapped in false local minima unless a sufficiently good starting point is provided. The latter is usually done by specifying a set of projection functions that approximate the sought WFs. In many cases, however, it is not easy to find a reasonable guess for the projection functions. This is particularly difficult when it comes to the construction of WFs from wide energy ranges of entangled bands, in systems with complex geometries or when the states are strongly hybridized. Although a recent study has proposed methods that are not based on projection \cite{Damle2015}, the MV algorithm is still the standard approach in the construction of MLWFs. A great improvement of the projection method has been made by \citet{Mustafa2015} who have introduced an algorithm that automatically constructs a set of optimized projection functions (OPFs) from a large pool of localized trial orbitals. For the valence bands of many materials such as SiO$_2$ and Cr$_2$O$_3$, the spread $\Omega$ of the initial guess obtained from the OPF method was shown to be just a few percent larger than the aimed global minimum \cite{Mustafa2015}.

Among the various ways of solving the KS equations of DFT, the full-potential linearized augmented planewave (LAPW) method, is considered to be the most accurate one. Highest numerical precision can be reached by selectively adding so-called local orbitals (LOs) to the LAPW basis set. The LOs are strongly localized atomic like functions. Hence, it is natural to employ the LOs in the construction of WFs within the LAPW+LO method. In this work, we combine the well established MV approach \cite{Marzari1997,Souza2001} with the more recent OPF technique \cite{Mustafa2015}. We employ the latter to automatically construct suitable initial guesses to MLWFs from a set of LOs. We demonstrate that this approach is capable to robustly construct MLWFs in a vast variety of materials without the need of manually providing a starting point. We apply the obtained WFs to study chemical bonding in a series of elemental and binary semiconductors. Further, electronic properties are calculated for more complex bulk and two-dimensional semiconductors as well as a hybrid organic-inorganic interface by the use of WFs based on (generalized) KS states and quasi-particle energies, respectively. We demonstrate that Wannier interpolation is capable to easily provide electronic energies with an accuracy in the meV~range over the entire Brillouin zone.

\section{\label{Methodology}Methodology}

\subsection{Theory of Wannier functions}

Here, we briefly discuss the basic steps in the construction of MLWFs and their application to interpolation. For an extensive overview over the MV approach, we refer to Ref.~\onlinecite{Marzari2012}.

Let $\psi_{n,\ve{k}}(\ve{r})$ be a set of single-particle Bloch wave functions describing a quantum-mechanical system as they may be obtained from a DFT calculation or any other method providing single (quasi-)particle eigenstates. In solids, the description of a quantum state in terms of Bloch functions is the natural choice, and the quantum numbers $n$ and $\ve{k}$ label an energy band and a wave vector in the first Brillouin zone (BZ), respectively. The Bloch formalism, however, is not the only way to describe quantum states in solids, and WFs provide an alternative representation. The transformation between a Bloch function $\phi_{n,\ve{k}}$ and a WF $w_{n,\ve{R}}$ reads \cite{Wannier1937}
\begin{equation}\label{eq-WFdef}
w_{n,\ve{R}}(\ve{r}) = \frac{1}{N_{\ve{k}}} \sum\limits_{\ve{k}} \e^{-\ci\ve{k}\cdot\ve{R}}\, \phi_{n,\ve{k}}(\ve{r})\;,
\end{equation}
where $\ve{R}$ is a real-space lattice vector labeling a unit cell within a supercell conjugate to the $\ve{k}$-point grid. Eq.~\eqref{eq-WFdef} holds for Bloch functions describing an isolated energy band. In solids, typically only deep-lying \mbox{(semi-)core} states form isolated bands. Therefore, it is desirable to generalize this transformation to a multitude of bands. To this end, we first consider an isolated group of energy bands, i.e. a group of $J$ bands that remains separated from all other bands by a finite energy gap throughout the BZ. The states $\psi_{n,\ve{k}}$ within this group span a subspace of the full space of solutions to the single-particle problem. Thus, they can be mixed according to some unitary transformation $U^{\ve{k}}$. The mixed states
\begin{equation}\label{eq-BlochMix}
\phi_{n,\ve{k}}(\ve{r}) = \sum\limits_{m=1}^J U^{\ve{k}}_{mn} \psi_{m,\ve{k}}(\ve{r})
\end{equation}
form an equally valid basis of the considered subspace and so do the WFs constructed according to Eq.~\eqref{eq-WFdef}. The unitary $J\times J$ matrix $U^{\ve{k}}$ reflects a generalization of the phase freedom of a single state and can be chosen freely. This freedom allows for the construction of WFs that are maximally localized according to some localization criterion. From another perspective, the matrices $U^\ve{k}$ define a gauge and are chosen such that the mixed states $\phi_{n,\ve{k}}$ are as smooth in $\ve{k}$ as possible, and consequently the Fourier transform in Eq.~\eqref{eq-WFdef} results in spatially well localized WFs. Although the valence bands in insulating or semi-conducting materials usually form such isolated groups, the conduction bands or the bands in metals often do not. In the case of such \textit{entangled} bands, first, at each $\ve{k}$-point a $J$-dimensional subspace
\begin{equation}\label{eq-DisStates}
\tilde{\psi}_{m,\ve{k}}(\ve{r}) = \sum\limits_{\mu=1}^{\mathcal{J}_{\ve{k}}} \mathcal{U}^{\ve{k}}_{\mu m}\psi_{\mu,\ve{k}}(\ve{r})
\end{equation}
has to be disentangled from the $\mathcal{J}_{\ve{k}}\geq J$ bands that fall inside a given (outer) energy window. This subspace is described by a rectangular $\mathcal{J}_{\ve{k}}\times J$ matrix $\mathcal{U}^{\ve{k}}$ which is semi-unitary (i.e. $\mathcal{U}^{\ve{k}\dagger} \mathcal{U}^{\ve{k}} = \openone_J$). Here, $J$ is the number of WFs one aims to construct from the bands inside an energy window of interest, and the $\psi_{\mu,\ve{k}}$ are single-particle wave functions whose eigenvalues fall inside that window. Furthermore, a second (inner) energy window can be introduced within which the states $\tilde{\psi}_{m,\ve{k}}$ in the disentangled subspace remain unchanged (i.e. $\mathcal{U}^{\ve{k}}_{\mu m} = \delta_{\mu m}$ for all states $\mu, m$ inside the inner window). Once the $J$-dimensional subspace is found, the construction of MLWFs is equivalent to the case of isolated bands with $\psi_{m,\ve{k}}$ replaced by $\tilde{\psi}_{m,\ve{k}}$ in Eq.~\eqref{eq-BlochMix}.

The MLWFs obtained from the above procedure form an excellent tight-binding basis which makes them suitable for an effective reciprocal-space interpolation in terms of a Slater--Koster interpolation \cite{Slater1954}. This Wannier interpolation scheme is based on the inversion of Eq.~\eqref{eq-WFdef} at an arbitrary point $\ve{q}$ in reciprocal space for which an interpolation is needed:
\begin{equation}\label{eq-AuxFromWan}
\phi_{m,\ve{q}}(\ve{r}) = \sum\limits_{\ve{R}} \e^{\ci\ve{q}\cdot\ve{R}}\, w_{m,\ve{0}}(\ve{r-R})\;.
\end{equation}
Eq.~\eqref{eq-AuxFromWan} describes the classical tight-binding approach and diagonalising the Hamiltonian matrix
\begin{equation}\label{eq-HamInt}
\mathcal{H}^{\ve{q}}_{mn} = \braket{\phi_{m,\ve{q}}|\ve{\hat{H}}|\phi_{n,\ve{q}}}
\end{equation}
gives rise to the single-particle eigenvalues $\epsilon^{\ve{q}}_n$ and eigenfunctions $\psi_{n,\ve{q}}$ at $\ve{q}$ expressed in terms of the auxiliary basis $\phi_{m,\ve{q}}$. The reason for the efficiency of this approach is that $\mathcal{H}^{\ve{q}}$ is easy to construct and typically much smaller than the Hamiltonian expanded in the original first-principles basis in which the states $\psi_{n,\ve{k}}$ are expressed. $\mathcal{H}^{\ve{q}}$ has the dimension $J$ (the number of bands under consideration) and therefore is easily diagonalized using standard linear-algebra routines.

\subsection{The (L)APW+LO method}
The approach described in detail below has been implemented into the full-potential all-electron code \exciting\;\cite{exciting} which is a realization of the (L)APW+LO method. This package implements DFT and many-body perturbation theory (MBPT). The latter is used to compute quasi-particle energies within the $G_0W_0$ approximation.

The APW method employs a partitioning of the unit cell into so called muffin-tin spheres (non-overlapping spheres centered at the nuclei) and an interstitial region (space between the muffin-tin spheres). The basis functions are planewaves in the interstitial region which are smoothly augmented into the muffin-tin spheres by atomic-like functions. The latter are expanded in terms of spherical harmonics around the nuclei. The corresponding radial functions $u_l(r;E_l)$ are solutions of the radial Schr\"odinger equation and parametrically depend on the energy $E_l$. In principle, the parameters $E_l$ have to be set to the band energies. In practice, however, these are not known a priori, and the basis itself would depend on the solution of the KS equations resulting in a non-linear eigenvalue equation. In order to linearize the eigenvalue problem, $E_l$ is set to a fixed value typically chosen to lie inside the respective band. In order to add more variational flexibility, the energy derivatives ${\dot{u}(r;E_l) = \partial u(r;E_l)/\partial E_l}$ can be added to the radial functions resulting in LAPWs.

This basis set can be further extended by the addition of so-called local orbitals (LOs). These functions are non-vanishing only inside one particular muffin-tin sphere at the atomic site $\ve{R}_{\alpha_\mathsf{L}}$, where they are given by
\begin{equation}\label{eq-LO}
\phi_\mathsf{L}(\ve{r}) = \left[\sum\limits_{o} a^\mathsf{L}_o\, u^{\alpha_\mathsf{L}}_{l_\mathsf{L},o}(|\ve{r}-\ve{R}_{\alpha_\mathsf{L}}|)\right]Y_{l_\mathsf{L} m_\mathsf{L}} (\widehat{\ve{r}-\ve{R}_{\alpha_\mathsf{L}}})\;.
\end{equation}
The coefficients $a^\mathsf{L}_o$ are chosen such that $\phi_\mathsf{L}$ is normalized and continuous at the muffin-tin boundary. The radial functions $u^{\alpha_\mathsf{L}}_{l_\mathsf{L},o}$ are solutions of the radial Schr\"odinger equation with a spherically symmetric potential inside the muffin-tin sphere, and the parameter $o$ denotes the linearization order (order of the derivative w.r.t. the energy parameter $E_l$). The addition of LOs results in a highly flexible and tunable basis set and allows for a smaller planewave cut-off.

Whenever high numerical precision is demanded, the full-potential (L)APW+LO method is considered the gold standard approach to first-principles calculations based on DFT and allows for the most precise numerical treatment of both ground state \cite{Gulans2018} and excited state properties \cite{Nabok2016}.

\subsection{Wannier functions from local orbitals}
The MV approach aims to find a set of unitary matrices $U^{\ve{k}}$ that minimizes the WF spread 
\begin{equation}\label{eq-Omega}
\Omega = \sum_n [\braket{w_{n,\ve{0}}|r^2|w_{n,\ve{0}}} - \braket{w_{n,\ve{0}}|\ve{r}|w_{n,\ve{0}}}^2]\;.
\end{equation}
In order to ensure a convergence of the iterative minimization of $\Omega$ and to minimize the risk of becoming trapped in false local minima a good starting point is indispensable. In our implementation, we avoid to manually provide suitable projection functions by the use of the OPF method \cite{Mustafa2015}. This method finds a guess to the MLWFs that is expanded as a linear combination of localized trial orbitals. Then, this guess is taken as the starting point for the MV approach. We construct OPFs from a pool of LOs from Eq.~\eqref{eq-LO} as they are part of the (L)APW+LO basis. The choice of LOs as trial functions is appealing for several reasons: i) They are already well localized by definition (non-zero only inside one muffin-tin sphere). ii) They fit any specific problem at hand since they depend on the actual potential in the system. iii) All integrals needed are already available within the (L)APW+LO method. In practice, we proceed as follows. For each atom, we successively add local orbitals with different angular character $Y_{l_\mathsf{L} m_\mathsf{L}}$ and a different number of nodes in the radial function to the pool of trial orbitals according to the aufbau principle. Then, if linear dependencies occur, we remove linearly dependent functions from the pool. Since the cost of the construction of OPFs strongly depends on the size of the pool, the amount of local orbitals can be further reduced to a specified number $N_\mathsf{L}$ by selecting the $N_\mathsf{L}$ local orbitals with the largest overlap with the states $\psi_{n,\ve{k}}$ in the considered subgroup (isolated bands) or energy window (entangled bands). 

\section{Results}

\subsection{Construction and chemical analysis}
\label{Construction}
\begin{figure}[h]
\includegraphics[width=\columnwidth]{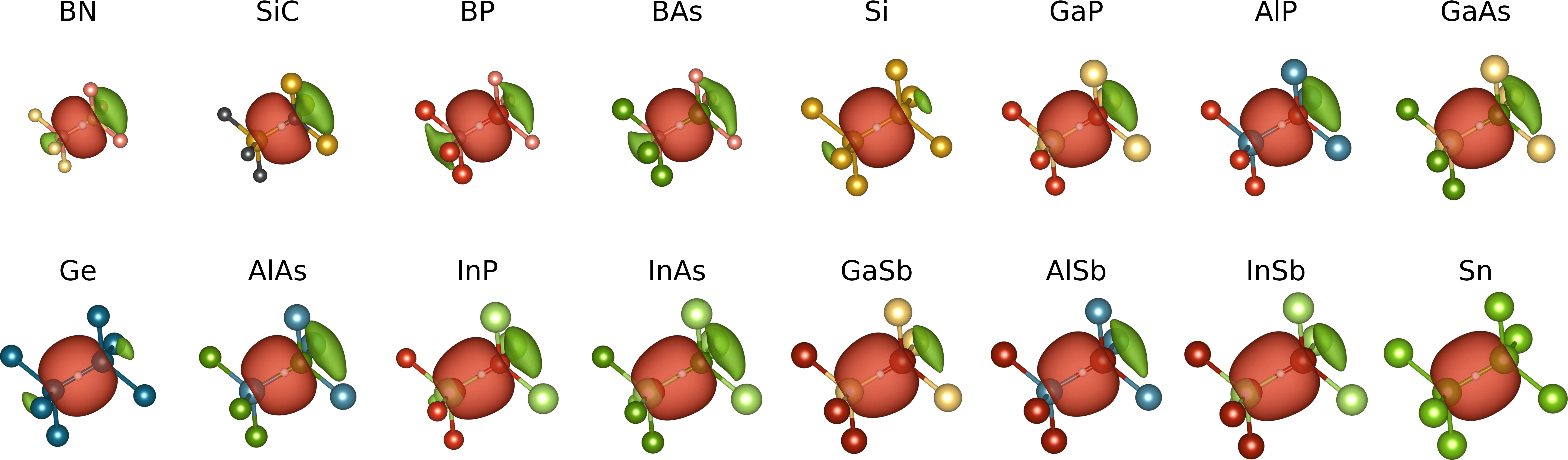}
\caption{\label{ionicityAll}MLWFs corresponding to the valence bands in 16 group IV and III--IV compounds in the diamond and zincblende structure. The white sphere on the central bond axis illustrates the WF center. All functions are real valued, and surfaces for the same positive (negative) iso-value are shown in red (green).}
\end{figure}

The localized nature of WFs and their formal exactness make the Wannier representation superior to the Bloch representation in terms of interpretation and chemical analysis. As an example for the chemical interpretation of MLWFs, we consider various group IV and III--V compounds crystallizing in the diamond or zinc-blende structure. All 16 considered materials (listed in Table~\ref{tab:ionicity}) are semiconductors and exhibit similar electronic properties. In particular, they form an isolated group of four distinct valence bands with hybridized $sp^3$-character for which we construct a set of four MLWFs. They transform into one another under symmetry operations, and each of them corresponds to one of the four tetrahedral bonds that each atom in these systems forms. The results are depicted in Fig.~\ref{ionicityAll}.
\begin{table}[htb]
\caption{\label{tab:ionicity}%
WF spreads $\Omega$ and shifts $\sigma$ of the WF centers calculated for 16 group IV and III-V compounds in the diamond (D) and zinc-blende (ZB) structure. The given lattice constants $a$ are adopted from Ref.~\onlinecite{Abu-Farsakh2007}.}
\begin{ruledtabular}
\begingroup
\setlength{\tabcolsep}{5pt} 
\renewcommand{\arraystretch}{1.1} 
\begin{tabular}{lcrrrrr}
& 
&
$a$ (\AA) &
\multicolumn{2}{c}{$\Omega$ (\AA$^2$)}&
\multicolumn{2}{c}{$\sigma$}\\
& & &
\textrm{Present} &
\textrm{Ref.~\onlinecite{Abu-Farsakh2007}} &
\textrm{Present} &
\textrm{ Ref.~\onlinecite{Abu-Farsakh2007}}\\
\hline
Si 	& D		&  5.431	&  8.200	&  8.232	&  0.000	&  0.000 \\
Ge 	& D		&  5.658	& 10.078	& 10.116	&  0.000	&  0.000 \\
Sn 	& D		&  6.490	& 13.752	& 13.801	&  0.000	&  0.000 \\
BP 	& ZB	&  4.540	&  5.532	&  5.479	&  0.034	&  0.032 \\
BAs	& ZB	&  4.777	&  6.207	&  6.211	&  0.048	&  0.052 \\
GaSb& ZB	&  6.100	& 11.390	& 11.527	&  0.146	&  0.154 \\
InSb& ZB	&  6.480	& 12.484	& 12.251	&  0.202	&  0.220 \\
GaP	& ZB	&  5.450	&  8.071	&  7.637	&  0.220	&  0.240 \\
GaAs& ZB	&  5.650	&  9.266	&  8.871	&  0.222	&  0.236 \\
AlSb& ZB	&  6.140	& 10.275	& 10.135	&  0.234	&  0.228 \\
InP	& ZB	&  5.870	&  9.370	&  8.492	&  0.274	&  0.308 \\
InAs& ZB	&  6.060	& 10.730	& 10.138	&  0.274	&  0.302 \\
SiC & ZB	&  4.360	&  4.741	&  4.651	&  0.302	&  0.308 \\
AlAs& ZB	&  5.660	&  8.197	&  8.090	&  0.310	&  0.310 \\
AlP	& ZB	&  5.460	&  7.250	&  7.146	&  0.312	&  0.314 \\
BN 	& ZB	&  3.620	&  2.857	&  2.820	&  0.314	&  0.316 \\
\end{tabular}
\endgroup
\end{ruledtabular}
\end{table}
\onecolumngrid

\begin{figure}[tb]
\includegraphics[width=\columnwidth]{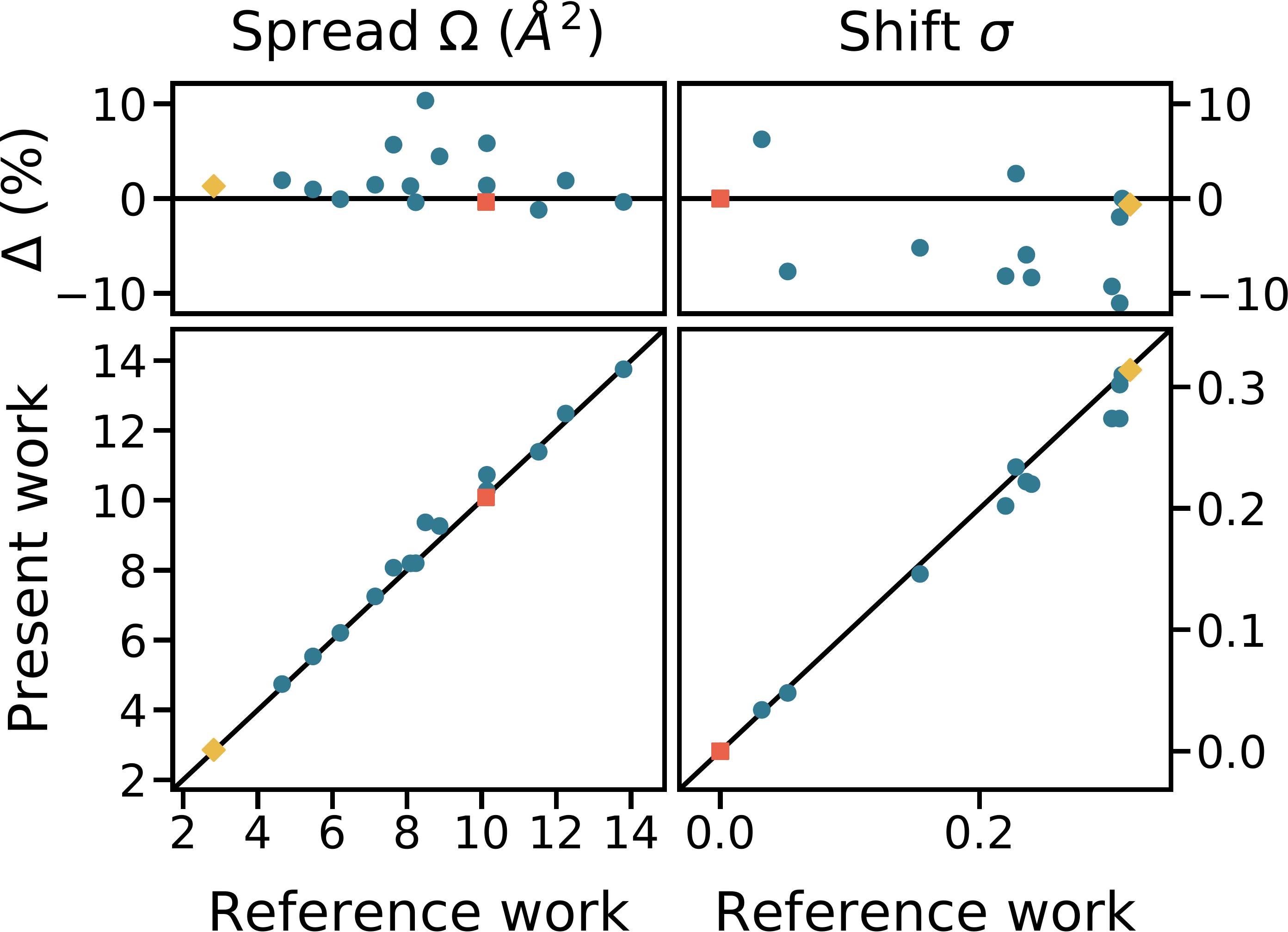}
\caption{\label{ionicity}{Spread $\Omega$ of the WFs (left) and shift $\sigma$ of their centers (right) for 16 group IV and III-V compounds. The results obtained within the present work are compared against Ref.~\onlinecite{Abu-Farsakh2007}. For both quantities, the relative deviation $\Delta = (A - A_{\rm ref})/A_{\rm ref}$ ($A=\Omega,\,\sigma$) is within 10\% (top). One purely covalent system (Ge) and the most ionic compound (c-BN) are highlighted by a red square and a yellow diamond, respectively.}}
\end{figure}

\newpage
\twocolumngrid
Indeed, the corresponding WFs have the character of a bonding $\sigma$-orbital, i.e. they are formed by a linear combination of the two $sp^3$-hybridized orbitals from both bonding atoms \cite{Marzari1997}. From visual inspection of these orbitals qualitative information about the bond character can be gained. For purely covalently bound systems (e.g. Ge) the WFs are symmetric and centered right in the middle of the bond while for more ionic bonds (e.g. c-BN) they are asymmetric and pushed towards the more electronegative atom (nitrogen in this example). Built upon this observation, \citet{Abu-Farsakh2007} proposed a first-principles parameter-free ionicity scale based on the position of the WF centers ${\braket{\ve{r}}_n = \braket{w_{n,\ve{0}}|\ve{r}|w_{n,\ve{0}}}}$. For 32 compounds of the type A$^N$B$^{8-N}$ ($N=1,\dots,4$), they defined the bond ionicity based on the parameter $\sigma$, describing the shift of the WF center away from the bond center ($\sigma=0$) towards the anion ($\sigma=1$). We use their findings to check our automated construction of MLWFs against an existing implementation for the 16 compounds studied here. As far as possible, the numerical parameters (lattice constants, $\ve{k}$-grids for obtaining the density and WFs, xc-type) are adopted from Ref~\onlinecite{Abu-Farsakh2007}. The results are shown in Fig.~\ref{ionicity}.
In all examples, our implementation finds the global minimum of the spread $\Omega$. It is worth noting that for this class of materials with bond-centered WFs the choice of LOs (which are strictly atom-centered and even vanish along the bond direction) as projection functions seems counterintuitive. Indeed, we find that the use of mere $s$- and $p$-like LOs as projection functions can result in a local minimum of the spread $\Omega$ corresponding to atom-centered WFs. However, this issue is fully resolved by employing suitable linear combinations of LOs obtained by the use of the OPF method.
In Table~\ref{tab:ionicity} we present both the spread $\Omega$ and the shift $\sigma$ for all 16 materials. With a relative deviation $\Delta$ of at most 10\%, both quantities are in good agreement with Ref.~\onlinecite{Abu-Farsakh2007} (top of Fig~\ref{ionicity}). We attribute these discrepancies to different approximations in the underlying first-principles calculation resulting in different densities and wave functions. While we employ a full-potential all-electron approach within the (L)APW+LO basis, in Ref.~\onlinecite{Abu-Farsakh2007} pseudopotentials and planewaves were used.

\subsection{Interpolation of energy eigenvalues}
\label{EnergyInterpolation}

\begin{figure}[htb]
\includegraphics[width=\columnwidth]{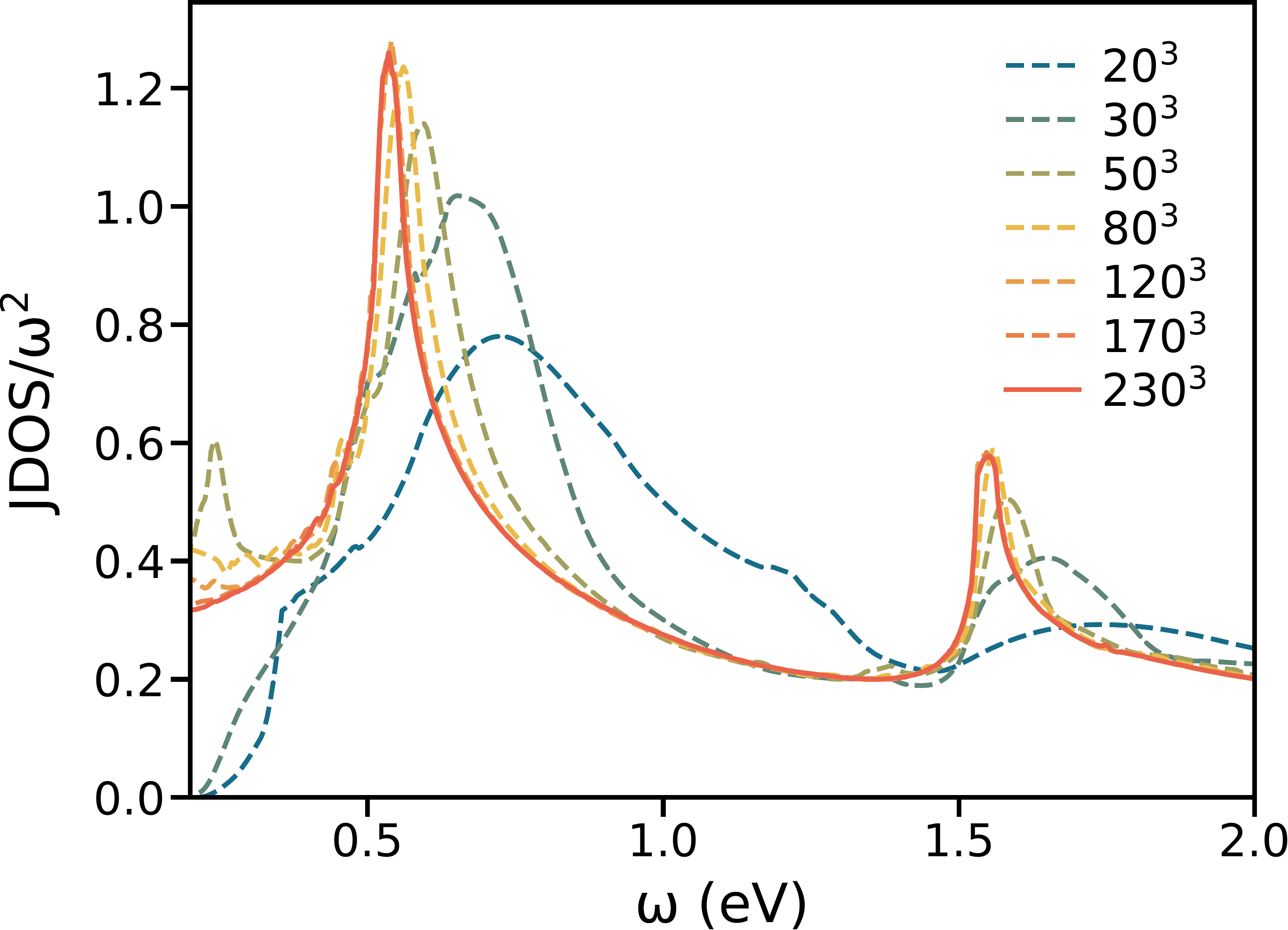}
\caption{\label{JDOS_Al}{Joint density of states for aluminum obtained from Wannier interpolation on different BZ-integration grids ranging from $20\times 20\times 20$ to $230\times 230\times 230$.}}
\end{figure}

The most obvious application of WFs is the interpolation of single-particle eigenenergies. For an arbitrary point $\ve{q}$ in reciprocal space, the corresponding energies $\epsilon_n^\ve{q}$ are given as the eigenvalues of the Hamiltonian matrix from Eq.~\eqref{eq-HamInt}. In practice, $\ve{q}$ is usually a point along a path connecting high-symmetry points in the BZ, when it comes to the calculation of band-structures, or a point on a grid which is denser than the original grid on which the first-principles calculation was carried out. Such dense grids are often used to approximate integrations over the BZ by a discrete sum over a finite set of points. One key quantity of interest that involves such a BZ-integration is the density of states (DOS). We use our implementation to investigate the joint DOS (JDOS) in aluminum. The JDOS is the phase space contribution to optical excitations and can be calculated as
\begin{equation}\label{eq-JDOS}
{\rm JDOS}(\omega) = \int\limits_{\rm BZ} \sum\limits_{o,u} \delta[ \epsilon_u(\ve{k}) - \epsilon_o(\ve{k}) - \omega]\,\ud\ve{k}\;,
\end{equation}
where $o$ and $u$ denote the occupied and unoccupied states for a given $\ve{k}$, respectively, and $\omega$ is the excitation energy. Note that the JDOS divided by $\omega^2$ is proportional to the independent-particle optical spectrum with constant transition matrix-elements. The spectrum of metals such as Al can be described well within the independent-particle picture since excitonic effects play a minor role due to the effective screening. Earlier calculations of optical spectra in Al showed that very dense integration grids containing several thousands irreducible $\ve{k}$-points are needed to obtain convergence of the spectra \cite{Lee1994,Ambrosch-Draxl2006}. In particular, also a strong dependence of the peak positions was observed \cite{Ambrosch-Draxl2006}. To investigate the influence of the BZ-grid on the JDOS in Al, we perform a DFT calculation within the generalized gradient approximation (GGA) using the PBE xc-functional \cite{Perdew1996} on a $12\times 12\times 12$ $\ve{k}$ mesh. From an outer (inner) energy window of -15\,eV to 80\,eV (-15\,eV to 40\,eV) 25 MLWFs are constructed using the disentanglement procedure. Hereby, the zero-energy point corresponds to the Fermi level. We interpolate the eigenvalues on different uniform integration grids by the use of MLWFs, and the improved tetrahedron method \cite{Kawamura2014} is employed to evaluate the integral in Eq.~\eqref{eq-JDOS}. The structural parameters used in this and all other calculations can be found in Table~\ref{tab:lattice}. The resulting JDOS is shown in Fig.~\ref{JDOS_Al}. We observe both a red shift and a significant sharpening of the two major peaks in the investigated energy region. Both peaks eventually converge at around 0.6\,eV and 1.6\,eV for $120^3$ and $80^3$ $\ve{k}$-points, respectively. We notice that it is more difficult to achieve convergence in the low-energy region. For energies below 0.4\,eV grids with more than $200^3$ uniformly spaced points are needed (solid red line). The position of the two peaks around 0.6\,eV and 1.6\,eV in the converged curve are in excellent agreement with earlier calculations of the JDOS \cite{Lee1994,Szmulowicz1981} as well as calculations \cite{Lee1994,Szmulowicz1981,Ambrosch-Draxl2006} and measurements \cite{Szmulowicz1981} of optical spectra. The red shift and sharpening with increasing grid densities was also found in calculations of optical spectra \cite{Ambrosch-Draxl2006}, where obviously the ratio of the peak heights differs from the optical spectra since transition probabilities are not taken into account in the JDOS. 

\subsection{Accuracy of energy interpolations}
\label{InterpolationAccuracy}

The claim that the WFs constructed according to Eq.~\eqref{eq-WFdef} form an equivalent description of the subspace spanned by the Bloch states $\psi_{n,\ve{k}}$ under consideration only holds rigorously for isolated groups and in the limit of an exact BZ integral. In practice, however, the BZ is sampled by a finite set of points. As a result, the computed WFs become periodic with respect to a supercell conjugate to the BZ grid. This can lead to a non-vanishing overlap between a WF and its periodic images in neighboring supercells and ruins the exactness of the tight-binding basis from Eq.~\eqref{eq-AuxFromWan} which is given in the limit of an exact BZ integral. In turn, the interpolated eigenvalues at some point $\ve{q}$ that does not belong to the original first-principles grid deviates from the exact solution. Note that \textit{exact} is meant within the limitations of the first-principles calculations, i.e. the interpolated energy deviates from the result one would obtain by directly performing the calculations at the point $\ve{q}$. MLWFs associated with isolated bands are reported to be exponentially localized \cite{He2001}. This claim was proved for insulators with time-reversal symmetry \cite{Brouder2007}. As a consequence, we expect the overlap with supercell images and thus the error in the interpolation to decay exponentially with increasing grid size. To investigate this behavior for the materials studied in this work, we proceed as follows. We consider a set of different grids $\lbrace\ve{k}\rbrace_1, \dots, \lbrace\ve{k}\rbrace_n$ (ordered with increasing grid density) for which we want to predict the accuracy of interpolated eigenenergies. First, we compute the self-consistent KS-potential and electron density on the densest grid under consideration $\lbrace\ve{k}\rbrace_n$. This self-consistent density serves as a starting point for further calculations. We use it to obtain the eigenvalues $\hat{\epsilon}_n^\ve{q}$ on a much denser interpolation grid $\lbrace\ve{q}\rbrace$ by a non self-consistent diagonalization of the KS-Hamiltonian. The dense interpolation grid is chosen to be shifted to ensure a sampling on inequivalent points. This set of energies $\hat{\epsilon}_n^\ve{q}$ forms the reference to which we compare the interpolated energies. Now, for each of the grids $\lbrace\ve{k}\rbrace_1, \dots, \lbrace\ve{k}\rbrace_{n-1}$ both wave functions and eigenenergies are calculated non self-consistently starting from the density obtained on the grid $\lbrace\ve{k}\rbrace_n$. Lastly, for all grids $\lbrace\ve{k}\rbrace_1, \dots, \lbrace\ve{k}\rbrace_n$ MLWFs are constructed and used to interpolate the eigenvalues onto the dense shifted interpolation grid $\lbrace\ve{q}\rbrace$. The interpolated energies are denoted by $\epsilon_n^\ve{q}$. For each grid, we compute the interpolation error as the root mean square deviation of the interpolated energies from the calculated reference energies:
\begin{equation}
\delta\epsilon_{\rm RMS} = \sqrt{\frac{1}{J N_\ve{q}}\sum\limits_{n,\ve{q}} (\epsilon_n^\ve{q} - \hat{\epsilon}_n^\ve{q})^2}\;.
\end{equation}
In order to compare BZ samplings for systems with different unit cell size and dimensionality, we introduce the linear $\ve{k}$-point density which is given by $(N_\ve{k}/V_{{\rm BZ},d})^{1/d}$, where $N_\ve{k}$ is the total number of non-reduced $\ve{k}$-points, $d$ is the dimensionality of the system, and $V_{{\rm BZ},d}$ is the volume of the corresponding $d$-dimensional BZ.

\begin{figure}[htb]
\includegraphics[width=\columnwidth]{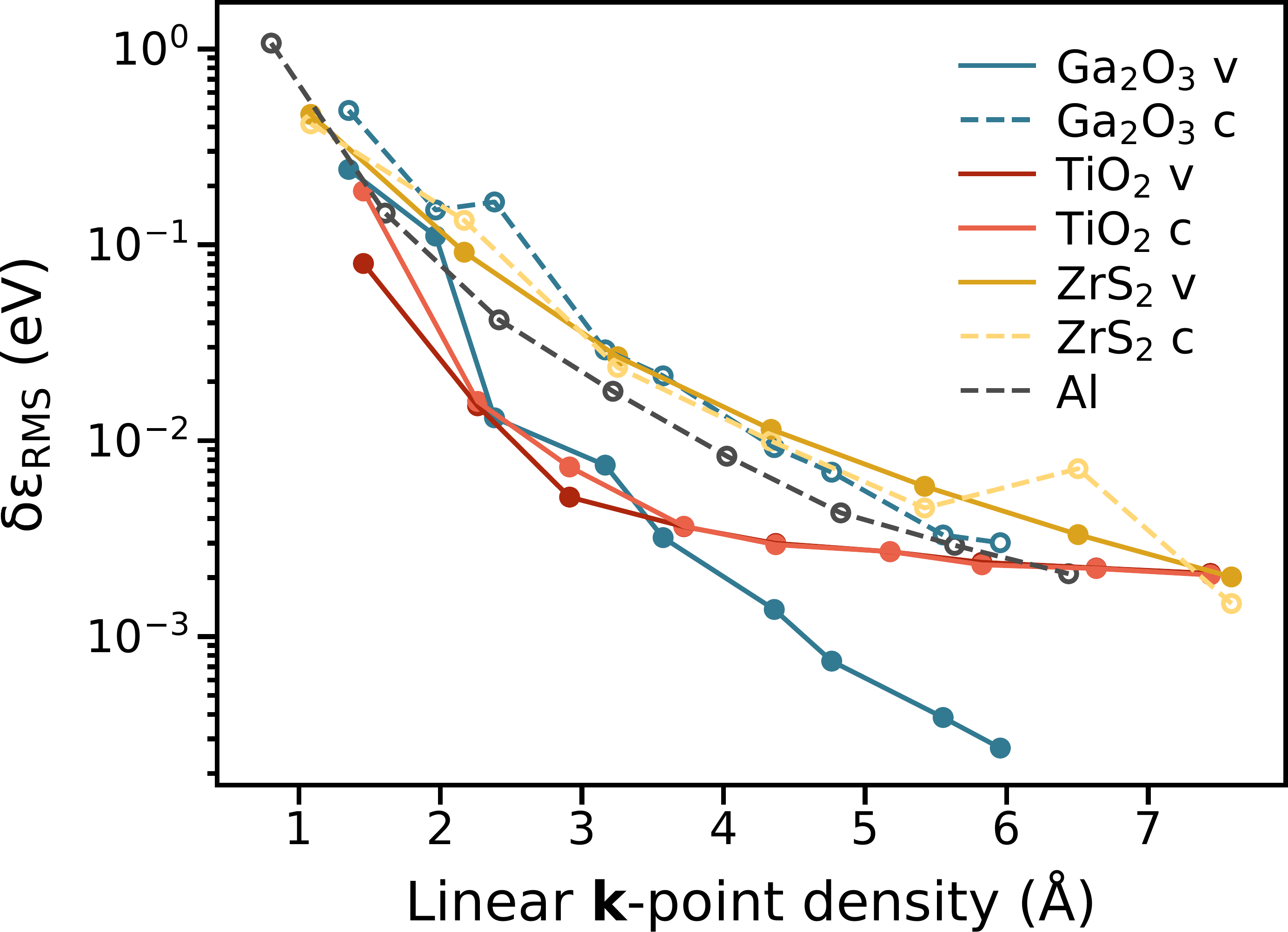}
\caption{\label{kconv}{Wannier interpolation error as a function of the $\ve{k}$-grid density. The filled (empty) circles mark the error of interpolated eigenvalues obtained from MLWFs representing isolated (entangled) bands in various systems for both valence (v) and conduction (c) bands. The lines serve as guides to the eye.}}
\end{figure}

We carry out DFT calculations for various materials using the PBE xc-functional and follow the procedure described above. The results are presented in Fig.~\ref{kconv}. The graphs indicate that an exponential decay of the interpolation error is an overall suitable assumption for most of the systems studied within this work. It is even found for the interpolation of entangled bands (empty circles, dashed lines) although there is no reason to assume an exponential localization of WFs obtained from the disentanglement procedure. The exponential decay is observed particularly well in the case of $\beta$-Ga$_2$O$_2$ for both the valence and the conduction bands. For TiO$_2$, however, the behavior differs considerably from a pure exponential decay. Similar investigations have been performed before for a set of isolated bands in lead and for entangled bands in lithium \cite{Yates2007}. There, the same behavior of a decreasing rate of decay for increasing grid densities (as it is clearly visible for TiO$_2$ in our calculations) was observed. Further, it was shown for 1D systems that the localization of energy matrix-elements follows a power law times an exponential \cite{He2001}. Such a model also fits well to our results obtained for 2- and 3-dimensional systems. For all systems studied, an interpolation accuracy in the meV~regime can be reached with manageable grid densities. Going to higher accuracies, however, will require higher grid densities than presented in Fig.~\ref{kconv} which may be feasible for KS-DFT eigenvalues but become rather cumbersome for the interpolation of generalized KS-eigenvalues obtained from hybrid xc-functionals or quasi-particle energies obtained from the $GW$ approach.

\subsection{Effective masses and band extrema}
\label{EffectiveMasses}

The accurate and inexpensive energy interpolation using WFs allows for a systematic search for band extrema. In semiconductors, the most interesting extremal points of the energy dispersion $\epsilon_n(\ve{k})$ typically are the highest occupied state (valence band maximum, VBM) and the lowest unoccupied state (conduction band minimum, CBm) determining the band gap and its type (direct or indirect). Finding their position is challenging when they are not located at a high-symmetry point in the BZ. In this case, they are usually not contained in the uniform BZ sampling employed in the DFT calculation. We use our implementation to determine the exact position of the VBM and CBm in $\beta$-Ga$_2$O$_3$ \cite{Furthmuller2016}, focusing on the effect of different xc-treatments and levels of theory. To this extent, the KS-equations are solved within the local-density approximation (LDA) parametrized by \citet{LDA_PW}, GGA using PBEsol \cite{Perdew2008}, and the non-local hybrid functional PBE0 with 25\% of Hartree--Fock exchange \cite{Ernzerhof1999}. Furthermore, quasi-particle self-energy corrections to the PBEsol eigenvalues are computed using the $G_0W_0$-approximation. The (generalized) KS calculations are carried out using $8\times 8\times 4$ $\ve{k}$-points in the full BZ. In the $G_0W_0$ calculation, a $4\times4\times4$ $\ve{k}$-mesh and all empty states are used following the prescription in Ref.~\onlinecite{Nabok2016}. The set of 18~valence bands is transformed into MLWFs using the algorithm for isolated groups. The spread $\Omega$ of the initial guess obtained from local orbitals using the OPF method is only 1\% larger than the global minimum for all xc-treatments. The WFs describing the conduction bands are obtained by the disentanglement procedure using an outer (inner) energy window of 30\,eV (20\,eV) above the Fermi level which was set to the middle of the band-gap.

\begin{figure}[htb]
\includegraphics[width=\columnwidth]{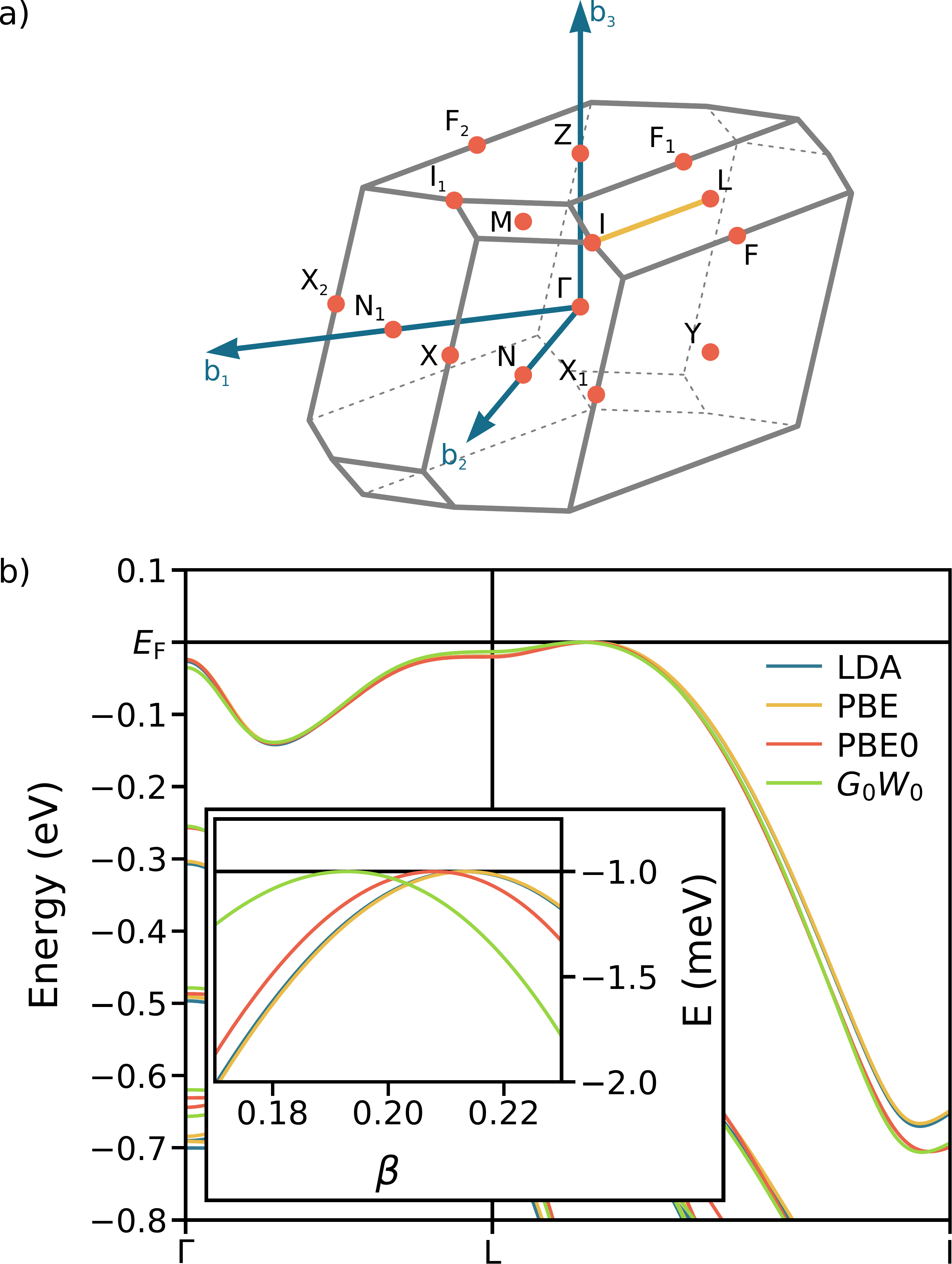}
\caption{\label{Ga2O3_VBM}{a) Brillouin zone of $\beta$-Ga$_2$O$_3$. The line on which the valence band maximum (VBM) is found is highlighted in yellow. b) Highest valence band in $\beta$-Ga$_2$O$_3$ along the high-symmetry lines $\Gamma$--L and L--I for different theoretical approaches. The Fermi level is set to the VBM. Inset: Region around the VBM. The parameter $\beta$ describes its position between the high-symmetry points L ($\beta = 0$) and I ($\beta = 1$).}}
\end{figure}

The CBm in $\beta$-Ga$_2$O$_3$ is known to be located at the zone-center $\Gamma$. This is confirmed by our calculations. The position of the VBM, however, is not at one of the high-symmetry points in the BZ. It was reported to be on the line connecting the high-symmetry points L and I \cite{Peelaers2015} (see Fig.~{\ref{Ga2O3_VBM}a}) which is in accordance with our findings. In Fig.~{\ref{Ga2O3_VBM}b}, we show the highest valence band and the position of the VBM for the different theoretical approaches. We find a weak dependence of the exact position of the VBM on the used xc-treatment. Although the position is nearly the same for LDA and PBEsol, it is slightly closer to L for the hybrid functional and the $G_0W_0$ calculation. The exact values are reported in Table~\ref{tab:meff_ana} by the parameter $\beta$ varying from 0 to 1 between the points L and I.

Further in Table~\ref{tab:meff_ana}, we present the resulting band gaps. A comparison of the $\Gamma$--$\Gamma$ gap with the experimental gap of about 4.9\,eV \cite{Orita2000,Janowitz2011} reveals that the non-local hybrid functional yields the best agreement with a direct gap of 5.0\,eV. As expected, the (semi-)local functionals LDA and PBEsol severely under estimate the gap. Also the quasi-particle gap of 4.5\,eV is underestimated. However, none of the theoretical values consider band renormalization effects due to electron-phonon interaction which makes a direct comparison with experimental results difficult. In all cases, the indirect gap is about 30\,meV smaller than the $\Gamma$--$\Gamma$ gap.

\begin{table}[hbt]
\caption{\label{tab:meff_ana}%
Position of the band extrema, effective masses (in units of $m_0$), and fundamental band gaps (in eV) in $\beta$-Ga$_2$O$_3$ for different xc-treatments determined analytically using Wannier interpolation. The parameter $\beta$ describes the position of the VBM along the line between the high-symmetry points L ($\beta = 0$) and I ($\beta = 1$).}
\begin{ruledtabular}
\begingroup
\setlength{\tabcolsep}{0pt} 
\renewcommand{\arraystretch}{1.1} 
\begin{tabular}{ldddd}
&
\multicolumn{1}{c}{\textrm{LDA}}&
\multicolumn{1}{c}{\textrm{PBE}}&
\multicolumn{1}{c}{\textrm{PBE0}}&
\multicolumn{1}{c}{$G_0W_0$@PBE}\\
\hline
VBM&&&\\
$\beta$	  &	0.2132	& 0.2136	& 0.2081	& 0.1953\\
$m^*_{xx}$&	2.94	& 2.95		& 2.97		& 3.20	\\
$m^*_{yy}$&	3.15	& 3.14		& 2.90		& 3.41	\\
$m^*_{zz}$&	4.30	& 4.39		& 4.73		& 3.02	\\
$m^*_{xz}$&	0.232	& 0.258		& 0.572		& 0.089 \\
\hline
CBm&&&\\
$m^*_{xx}$&	0.238	& 0.234		& 0.275		& 0.294	\\
$m^*_{yy}$&	0.263	& 0.263		& 0.280		& 0.333	\\
$m^*_{zz}$&	0.253	& 0.251		& 0.273		& 0.280	\\
\hline
$E_{\rm g}$ (eV)&&&\\
indirect  			& 2.271	& 2.290		& 5.009		& 4.490	\\
$\Gamma$--$\Gamma$	& 2.297	& 2.314		& 5.033		& 4.525	\\
\end{tabular}
\endgroup
\end{ruledtabular}
\end{table}

The simple form of the single-particle wave functions expressed in the WF tight-binding basis (see Eq.~\ref{eq-AuxFromWan}) allows for an analytic expression of $\ve{k}$-space derivatives since the dependence on the wave vector only comes from the exponential factor while the WFs themselves are $\ve{k}$-independent. Thus, the use of finite differences or numerical fitting methods (which are usually used to calculate derivatives) can be avoided. This analytical approach allows for the direct calculation of the particle group-velocity
\begin{equation}\label{eq-vel}
\ve{v}_n(\ve{k}) = \nabla_{\ve{k}} \epsilon_n(\ve{k})
\end{equation}
and the effective-mass tensor
\begin{equation}\label{eq-mass}
\ve{m}^*_n(\ve{k}) = \left[ \nabla_{\ve{k}} \nabla_{\ve{k}}^{\rm T} \epsilon_n(\ve{k}) \right]^{-1}\;.
\end{equation}
Note that atomic units are used in Eqs.~\eqref{eq-vel} and \eqref{eq-mass}, and $\nabla_{\ve{k}}$ is a column vector. We follow the derivations by \citet{Yates2007} in order to evaluate the analytic expression of the first and second $\ve{k}$-derivative of the band dispersion in $\beta$-Ga$_2$O$_3$ to determine the effective masses at the CBm and VBM. The results are given in Table~\ref{tab:meff_ana}. We find $\ve{m}^*_{\rm CBm}$ to be almost diagonal and isotropic. The electron effective mass varies from 0.237 to 0.333 electron rest-masses depending on the direction and the xc-treatment. Again, there are no noticeable differences between LDA and PBEsol. For the hybrid functional PBE0 the CBm is more isotropic compared to LDA and PBEsol, and the electrons are slightly heavier with effective masses between 0.273\,$m_0$ and 0.280\,$m_0$. These values are in perfect agreement with the (almost isotropic) value of 0.281\,$m_0$ previously reported for the hybrid functional HSE06 \cite{Varley2010}. The results for LDA and PBEsol are in good agreement with values of around 0.23\,$m_0$ that were previously obtained for LDA \cite{Yamaguchi2004}. For the quasi-particles we find even higher effective masses between 0.280\,$m_0$ and 0.333\,$m_0$.

At the VBM, the effective-mass tensor takes the following form:
\begin{equation}
\ve{m}^*_{\rm VBM} = \begin{pmatrix}
m^*_{xx} & 0 & m^*_{xz} \\
0 & m^*_{yy} & 0 \\
m^*_{xz} & 0 & m^*_{zz}
\end{pmatrix}\;,
\end{equation}
where the $m^*_{xy}$ and $m^*_{yz}$ components do not vanish completely but are about three orders of magnitude smaller than the diagonal components and therefore neglected. According to our calculations, the VBM is more anisotropic. For LDA, PBEsol, and PBE0, we obtain similar hole effective masses in the $x$- and $y$-direction of around 3\,$m_0$ and values between 4.3\,$m_0$ (LDA) and 4.7\,$m_0$ (PBE0) in the $z$-direction. The quasi-particle calculation differs noticeably from the other three approaches and suggests heavier holes in the $x$- and $y$-direction and lighter holes in the $z$-direction. Overall, our results are comparable with those of Ref.~\onlinecite{Yamaguchi2004} for the $y$- and $z$-direction but differ noticeably in the $x$-direction for which Ref.~\onlinecite{Yamaguchi2004} reported a hole effective mass of $m^*_{xx}=6.14\,m_0$ which is about twice the value we find. However, both the exact position of the VBM and the band curvature are difficult to determine accurately due to the very low dispersion in the valence band top region and the occurrence of multiple maxima that differ only little in energy. For instance, there is another maximum at $\Gamma$ only 30\,meV below the VBM see Fig.~\ref{Ga2O3_VBM}). We are not aware of any reports on experimental hole effective masses in $\beta$-Ga$_2$O$_3$ to compare with.

\begin{figure}[htb]
\includegraphics[width=\columnwidth]{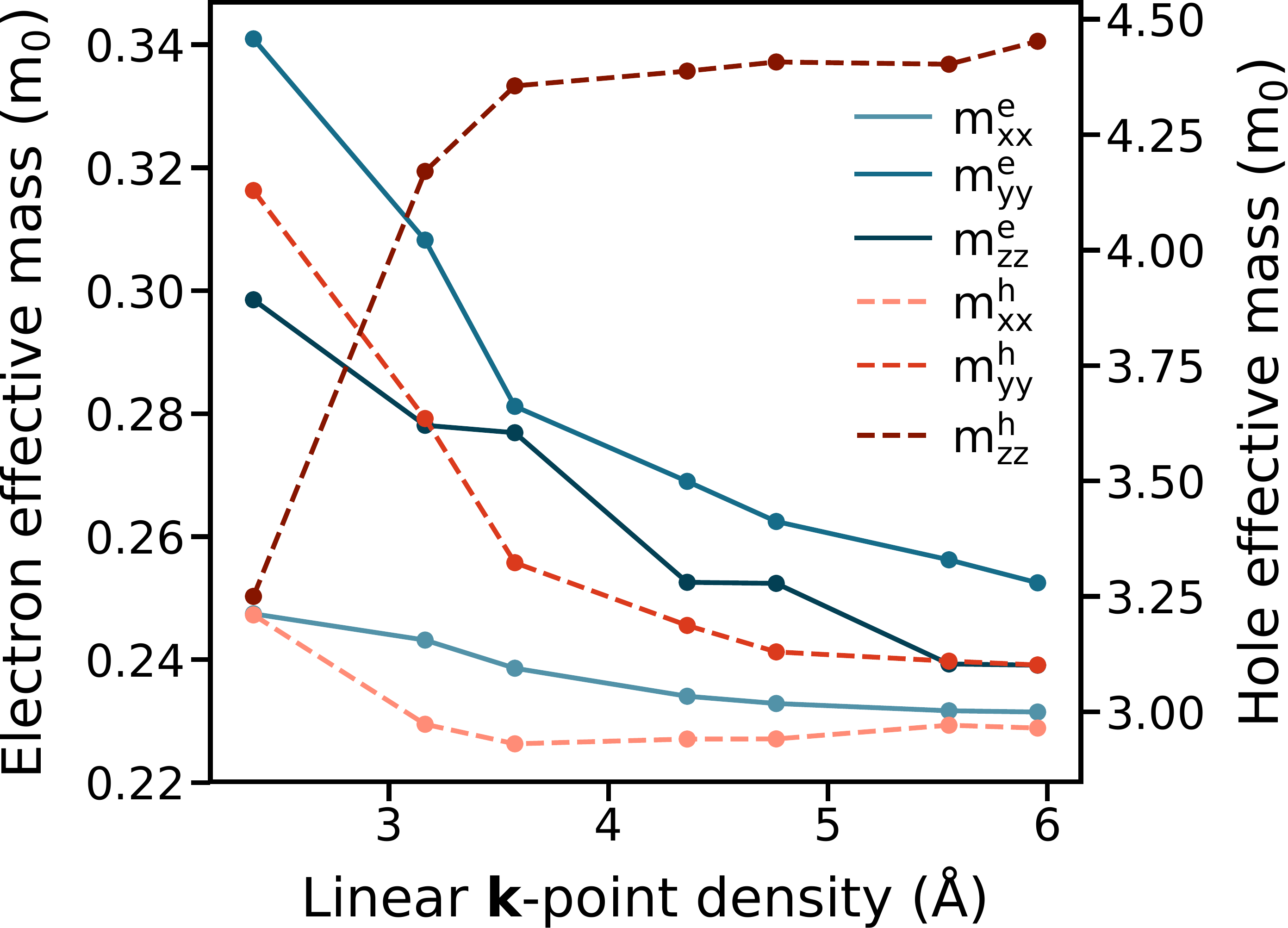}
\caption{\label{kconv_mass}{Diagonal effective masses for electrons (blue, solid lines) and holes (red, dashed lines) in $\beta$-Ga$_2$O$_3$ obtained from Wannier interpolation on top of PBE as a function of the first-principles $\ve{k}$-grid density.}}
\end{figure}

In order to estimate the accuracy of the determined effective masses, we perform a similar convergence test as it is done in Section \ref{InterpolationAccuracy} for the predicted energies. In Fig.~\ref{kconv_mass} we present the diagonal entries of the effective mass tensor for electrons at the CBm (blue, solid lines) and holes at the VBM (red, dashed lines) obtained from the analytic approach starting from PBEsol calculations on different $\ve{k}$-grids. The values presented in Table~\ref{tab:meff_ana} (with the exception of $G_0W_0$) are obtained on a grid corresponding to a linear $\ve{k}$-point density of about 4.8\,\AA. Fig.~\ref{kconv_mass} shows that for this grid density the hole effective masses are almost converged and we estimate an uncertainty of about 0.1\,$m_0$ ($\approx 3$\%). In contrast, the noticeably smaller electron effective masses are much harder to predict accurately. They are not yet fully converged in the studied range of $\ve{k}$-point densities and thus we estimate a larger relative uncertainty for the numbers in Table~\ref{tab:meff_ana} of about 0.02\,$m_0$ ($\approx 10$\%).

\subsection{Interpolation of wave functions}
\label{WavefunctionInterpolation}

The diagonalization of the Wannier-interpolated Hamiltonian $\mathcal{H}^\ve{q}_{mn}$ gives also rise to the interpolated wave functions. They are expressed in the form
\begin{equation}
\psi_{n,\ve{q}}(\ve{r}) = \sum\limits_m V^\ve{q}_{mn} \phi_{m,\ve{q}}(\ve{r})\;,
\end{equation}
where $V^\ve{q}_{:n}$ is the eigenvector of $\mathcal{H}^\ve{q}$ corresponding to the eigenvalue $\epsilon^\ve{q}_n$, and $\phi_{m,\ve{q}}$ is defined by Eq.~\eqref{eq-AuxFromWan}. The analysis of these wave functions offers deeper physical and chemical insights. To this extent, we decompose $\psi_{n,\ve{q}}$ in particular atomic states by an expansion in a series of spherical harmonics $Y_{lm}$ times radial functions $\varphi^\alpha_{n,\ve{q},lm}$ inside the individual muffin-tin spheres $\alpha$:
\begin{equation}
\psi_{n,\ve{q}}^\alpha(\ve{r}) = \sum\limits_l \sum\limits_{m=-l}^l \varphi^\alpha_{n,\ve{q},lm}(|\ve{r-R}_\alpha|) Y_{lm}(\widehat{\ve{r-R}_\alpha})\;.
\end{equation}
Within the (L)APW+LO basis, this expansion is straightforward. Now, we calculate the contribution of the state $\psi_{n,\ve{q}}$ to the number of electrons inside the muffin-tin sphere~$\alpha$ with radius $R_\alpha$ by integrating the partial density $\rho_{n,\ve{q}}(\ve{r}) = |\psi_{n,\ve{q}}(\ve{r})|^2$:
\begin{equation}\label{eq-bandchar}
\begin{aligned}
\int\limits_{{\rm MT}_\alpha} \rho_{n,\ve{q}}(\ve{r})\,\ud\ve{r} &= \sum\limits_l \sum\limits_{m=-l}^l \int\limits_0^{R_\alpha} r^2 |\varphi^\alpha_{n,\ve{q},lm}(r)|^2\,\ud r \\
&= \sum\limits_l b^{\alpha, l}_{n,\ve{q}}\,.
\end{aligned}
\end{equation}
The second line of Eq.~\eqref{eq-bandchar} defines the band character $b^{\alpha,l}_{n,\ve{q}}$ which is interpreted as the contribution of electrons with angular character $l$ and wave vector $\ve{q}$ inside the muffin-tin sphere $\alpha$ to the $n$-th energy band.

\begin{figure}[htb]
\includegraphics[width=\columnwidth]{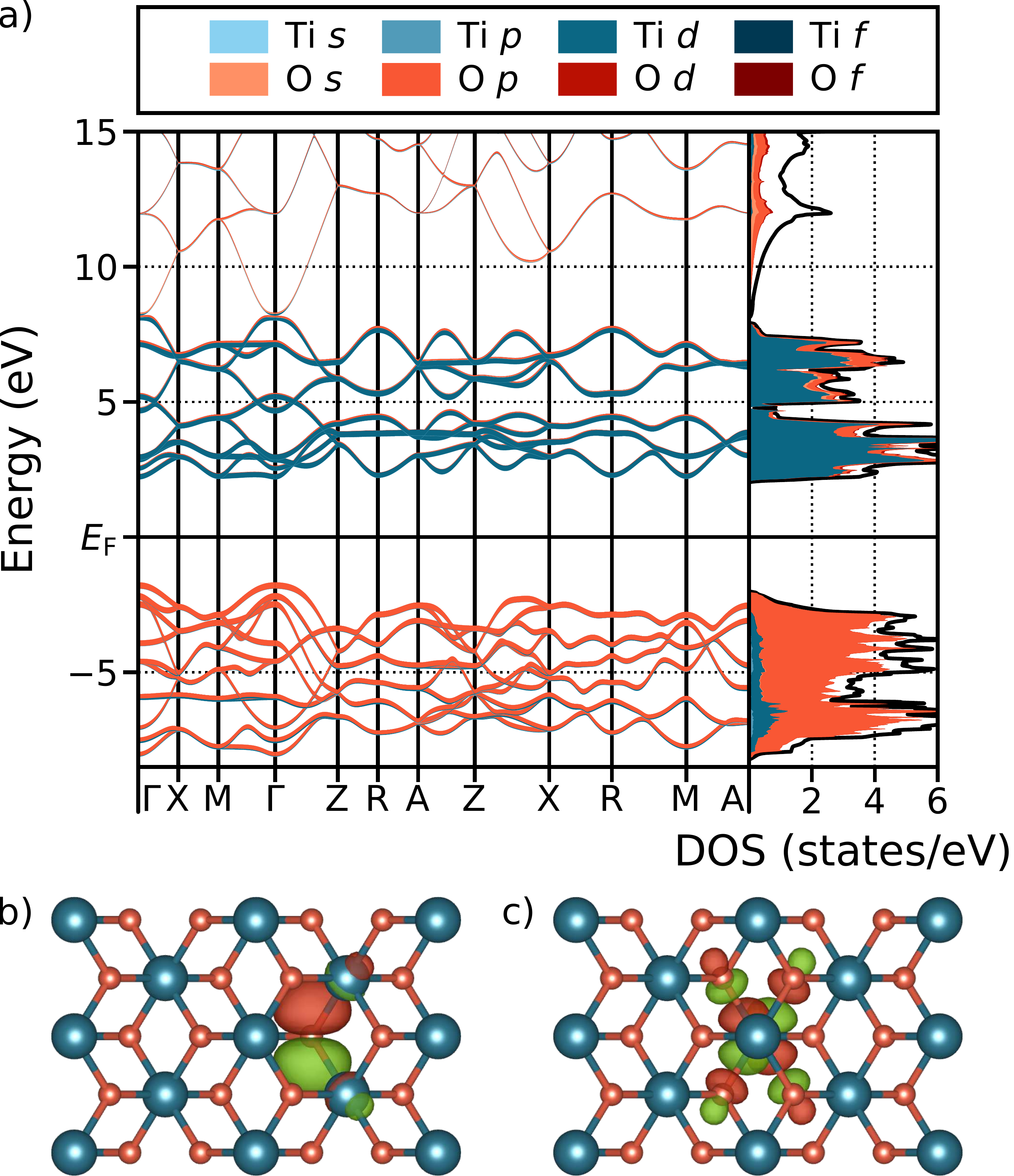}
\caption{\label{PBNDDOS_TiO2}{Wannier interpolated band-structure and DOS (a) for TiO$_2$ in the rutile structure calculated using PBE0. The Fermi level is set to the middle of the gap. The different shades of colors display the individual contributions of the wave functions at titanium (blue) and oxygen (red) atoms with different angular character ($l$). MLWFs corresponding the valence (b) and lowest conduction (c) bands. Note that the MLWFs are real-valued. Positive (negative) iso-surfaces are displayed in red (green).}}
\end{figure}

\begin{figure}[htb]
\includegraphics[width=\columnwidth]{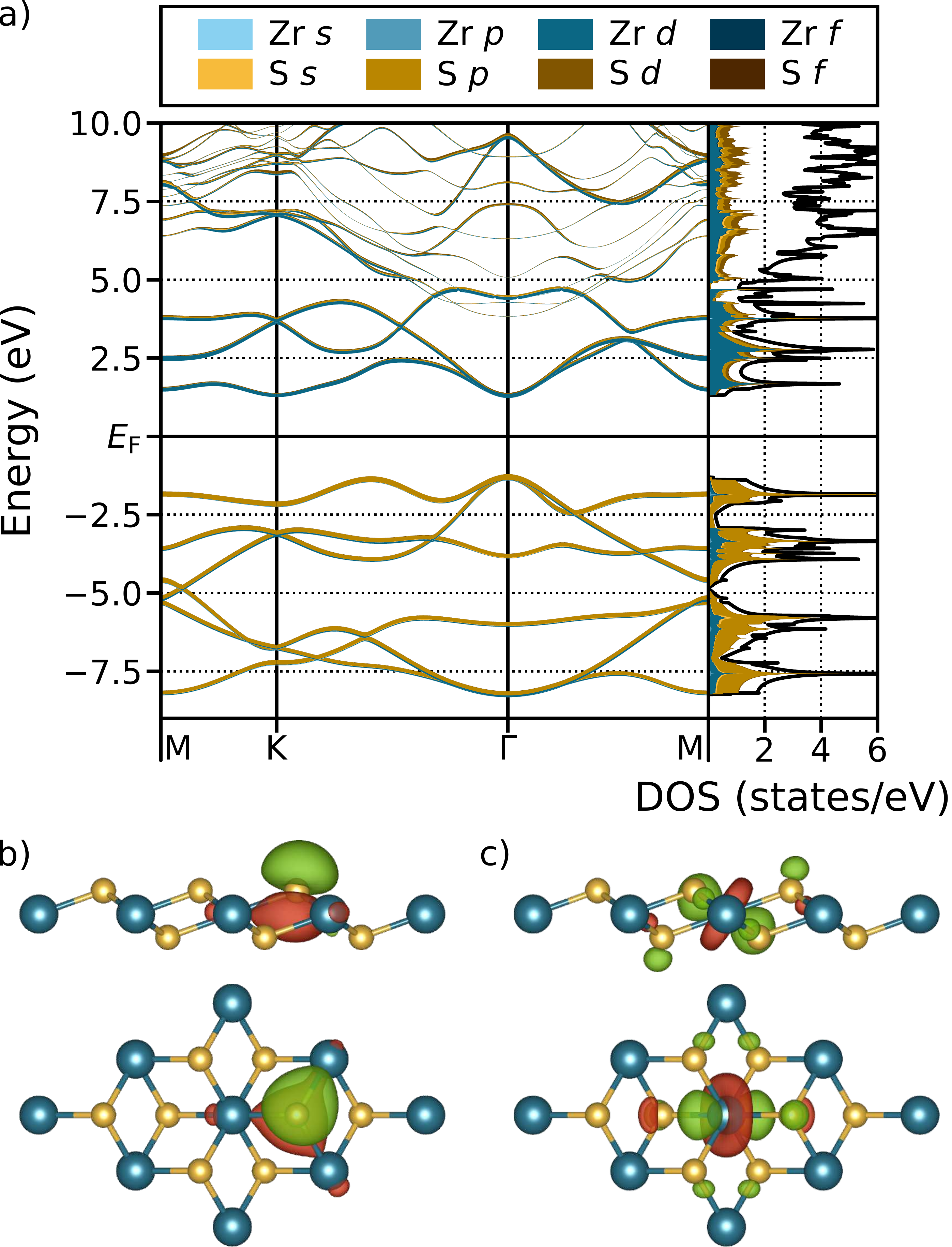}
\caption{\label{PBNDDOS_ZrS2}{Same as Fig.~\ref{PBNDDOS_TiO2} for a ZrS$_2$ monolayer calculated using the $G_0W_0$ approximation on top of PBE. The upper (lower) illustration of the MLWFs show the side (top) view.}}
\end{figure}

We interpolate the band character for TiO$_2$ in the rutile structure and for a monolayer of the 2D material ZrS$_2$. The calculation of TiO$_2$ is carried out using the hybrid xc-functional PBE0 and a $6\times 6\times 9$ $\ve{k}$-point grid. The 12 valence bands and the 10 lowest conduction bands are transformed into MLWFs separately using the algorithm for isolated bands. Again, the spread $\Omega$ of the initial guess is only 1\% and 2\% off the global minimum for the two groups, respectively. For the higher conduction bands, 148~WFs are disentangled using an outer (inner) energy window of 8\,eV to 130\,eV (8\,eV to 76\,eV). In the case of ZrS$_2$, quasi-particle energies are calculated within the $G_0W_0$ approximation on top of PBE for $8\times 8\times 1$ $\ve{k}$-points. The six valence bands are treated as an isolated group. We disentangle the three Zr $d$-like bands which intersect with higher energy conduction bands around the $\Gamma$-point from the energy window between 0\,eV and 4.75\,eV. For both the valence bands and the three disentangled conduction bands, the initial guess is 2\% larger than the global minimum. The remaining conduction bands are represented by 27~WFs disentangled from an outer (inner) energy window of 3.75\,eV to 20\,eV (4.75\,eV to 10\,eV). In the top panels of Figs.~\ref{PBNDDOS_TiO2} and \ref{PBNDDOS_ZrS2}, we present the interpolated band-structure and DOS for TiO$_2$ and ZrS$_2$, respectively. For obtaining the DOS, the energies and the band characters are interpolated on a grid of $60\times 60\times 90$ and $300\times 300\times 1$ points in the BZ for TiO$_2$ and ZrS$_2$, respectively. The bands and the DOS are colored according to the band character, i.e. the contribution of electrons from different atom species and with different angular character. Since the band character does not account for contributions from the interstitial region, the sum of the projected DOS (colored area) differs from the total DOS (black solid line). In the case of TiO$_2$, the 12 valence bands almost entirely originate from oxygen $p$-like states. The 12 symmetry-equivalent WFs corresponding to this group of bands (one illustrated in Fig.~\ref{PBNDDOS_TiO2}b) clearly reflect this character. The same holds for the isolated group of the 10 lowest conduction bands which exhibit dominantly titanium $d$ character with some admixture of oxygen $p$-like states. Again, this is clearly reflected in the corresponding Wannier functions (Fig.~\ref{PBNDDOS_TiO2}c). A similar behavior can be found in ZrS$_2$. The valence bands show a strong sulphur $p$ character since the corresponding Wannier functions (Fig.~\ref{PBNDDOS_ZrS2}b) are almost purely $p$-like and centered at sulphur atoms. The Wannier functions corresponding the three zirconium $d$-like bands in the lower conduction band region (Fig.~\ref{PBNDDOS_ZrS2}c) clearly reflect the dominant Zr $d$-character but also show contributions from sulphur $p$-like states.

\begin{figure}[htb]
\includegraphics[width=\columnwidth]{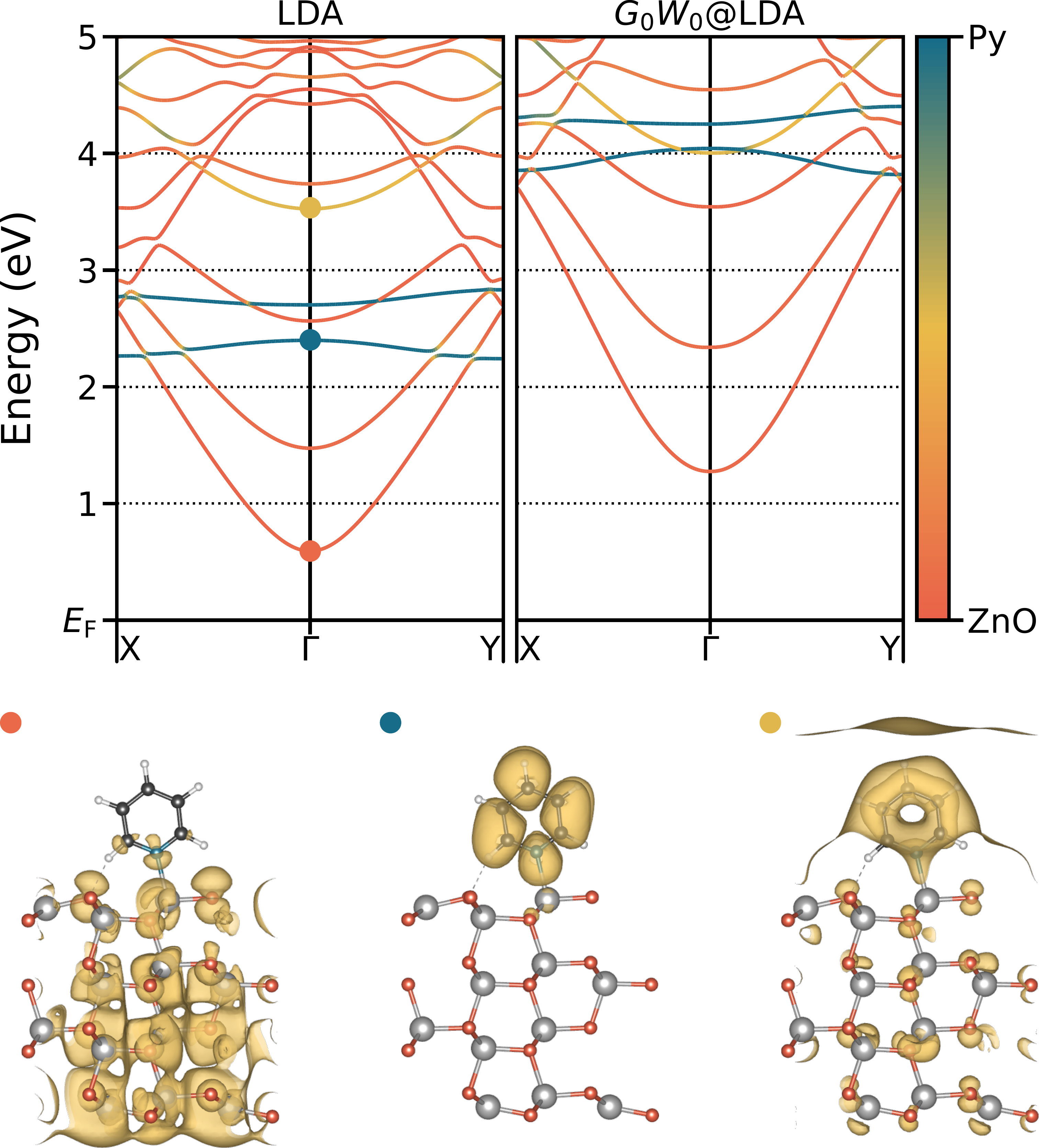}
\caption{\label{PBND_PyZnO}{Energy dispersion for the lowest KS and quasi-particle conduction bands in a hybrid inorganic-organic system (top left and top right, respectively). The bands are colored according to their origin. Bands attributed to the inorganic ZnO slab and the organic pyridine molecule are drawn in red and blue, respectively. The KS wave function for a hybridized state (yellow dot) as well as for states originating from ZnO (red dot) and pyridine (blue dot) are illustrated in the lower part.}}
\end{figure}

Within the $G_0W_0$ approximation, a self-energy correction to the KS eigenvalues is calculated in order to obtain the quasi-particle energies. Often, these corrections (obtained on a uniform $\ve{k}$-grid) are used to deduce a rigid scissors shift from which the band-structure is then obtained. This approach, however, is not always justified, like for instance in hybrid inorganic-organic systems. The prototypical compound shown here \cite{Turkina2019} consists of pyridine molecules chemisorbed on the ($10\overline{1}0$) surface of a ZnO slab with 43 atoms in the unit cell (see bottom panel in Fig.~\ref{PBND_PyZnO}). The quasi-particle energies are computed on $4\times4\times1$ $\ve{k}$-points corresponding to a linear $\ve{k}$-point density of 3.9\,\AA. From an outer (inner) window of 13.6\,eV (8.2\,eV) above the Fermi level 60~WFs are disentangled to compute the quasi-particle band-structure and compare it to the KS band-structure (Fig.~\ref{PBND_PyZnO}). Using the band character, we can attribute the individual energy bands to the constituents of the system. Bands displayed in blue are attributed to the organic molecule while red bands originate from the inorganic ZnO slab. Hybridized bands are colored in shades of green, yellow and orange. In the bottom of Fig.~\ref{PBND_PyZnO}, KS orbitals at $\Gamma$ are shown, attributed to ZnO (red), pyridine (blue), and a hybridized state (yellow), respectively. It is evident that the quasi-particle self-energy correction has significantly different effects on the individual energy bands depending on their origin. While all conduction bands experience a general shift towards higher energies, the two flat molecular bands (blue) are subject to a much stronger upwards shift with respect to the four parabolic ZnO bands (red). In contrast, the strongly hybridized band (yellow) is slightly shifted downwards with respect to ZnO bands.

\section{Conclusions}
We have presented an implementation of MLWFs within the (L)APW+LO method. By combining the well established algorithm developed by \citet*{Souza2001} with the more recently presented OPF technique \cite{Mustafa2015}, we are able to robustly construct MLWFs for various classes of materials without the need of projection functions being selected by the user. We use LOs as projection functions within the (L)APW+LO method. It is appealing due to its simplicity although they are strictly atom-centered and vanishing in the interstitial region. This lack of flexibility can be overcome reliably by the use of the OPF approach. By automatically and systematically adding and selecting LOs from the pool of projection functions, we are able to calculate MLWFs for both isolated and entangled bands in 2D and bulk semiconductors with small and medium sized unit cells, in metals as well as in complex hybrid systems containing an inorganic semiconductor and organic molecules.

This procedure gives access to accurate band structures and DOS based on more sophisticated methods such as generalized hybrid KS-DFT or quasi-particle calculations which otherwise would not be available due to the immense computational cost these methods come with. The same holds for other quantities that can be derived from the band structure directly such as band gaps, group velocities, and effective masses. According to our findings, a linear density of about {4 $\ve{k}$-points per \AA$^{-1}$} in reciprocal space in the underlying calculation suffices to predict electronic energies at an arbitrary point with an accuracy in the meV-range. A deeper analysis of the interpolated wave function gives access to the band character and allows for a detailed interpretation of band structures and DOS. The results are in excellent agreement with calculations carried out in the original basis indicating that not just eigenenergies but also wave functions can be predicted accurately.

Future applications may involve MLWFs as basis functions in excited state calculations using MBPT which often come with high effort simultaneously requiring dense $\ve{k}$-grids. A reduction of the basis size and the simple access to wave functions and energies at arbitrary points in reciprocal space may help to reduce the computational cost of these approaches retaining the high precision of the (L)APW+LO method.

\begin{acknowledgments}
This work was partially performed in the framework of GraFOx, a Leibniz-ScienceCampus supported by the Leibniz association. Parts of this work were funded by the Deutsche Forschungsgemeinschaft (DFG, German Research Foundation) - Projektnummer 182087777 - SFB 951. All input and output files can be downloaded from the NOMAD Repository, DOI: 10.17172/NOMAD/2019.08.28-1. The LDA and $G_0W_0$ calculations of the Py@ZnO interface underlying our investigations were performed by Olga Turkina, the $G_0W_0$ calculation of $\beta$-Ga$_2$O$_3$ by Dmitrii Nabok. We thank them for providing the data.

\end{acknowledgments}

\appendix

\onecolumngrid

\begin{table}[htb]
\caption{\label{tab:lattice}%
Structural parameters for all materials investigated in Sections \ref{EnergyInterpolation} to \ref{WavefunctionInterpolation}.}
\begin{ruledtabular}
\begingroup
\setlength{\tabcolsep}{0pt} 
\renewcommand{\arraystretch}{1.1} 
\begin{tabular}{lddddddlll}
\multicolumn{1}{c}{\textrm{Compound}}&
\multicolumn{6}{c}{\textrm{Lattice parameters}}&
\multicolumn{1}{c}{\textrm{Lattice}}&
\multicolumn{1}{c}{\textrm{Space}}&
\multicolumn{1}{c}{\textrm{Atoms in unit cell}}\\
& \multicolumn{1}{c}{a\textrm{\,(\AA)}} & \multicolumn{1}{c}{b\textrm{\,(\AA)}} & \multicolumn{1}{c}{c\textrm{\,(\AA)}} & \alpha\,(^\circ) & \beta\,(^\circ) & \gamma\,(^\circ) & \multicolumn{1}{c}{\textrm{type}}& \multicolumn{1}{c}{\textrm{group}} & \\
\hline
Al & 2.838 & 2.838 & 2.838 & 60 & 60 & 60 & cubic & Fm-3m & 1Al \\
Ga$_2$O$_3$ & 6.302 & 6.302 & 5.807 & 76.6 & 103.4 & 152.1 & monoclinic & C2/m & 4Ga, 6O \\
TiO$_2$ & 4.638 & 4.638 & 2.969 & 90 & 90 & 90 & tetragonal & P4(2)/mnm & 2Ti, 4O \\
ZrS$_2$ & 3.66 & 3.66 & 20.128 & 90 & 90 & 60 & hexagonal & P3m1 & 1Zr, 2S \\
Py@ZnO & 6.310 & 6.310 & 23.284 & 90 & 90 & 112.3 & triclinic & P1 & 5C, 5H, 1N, 16Zn, 16O

\end{tabular}
\endgroup
\end{ruledtabular}
\end{table}
\twocolumngrid

\bibliography{Tillack_bibliography}

\end{document}